\documentclass[12pt]{article}
  \usepackage{amsfonts}
  \usepackage{amsmath}
\usepackage{amssymb}
\usepackage{amscd}
\usepackage[dvips]{graphicx}
\begin{document}
~~
\bigskip
\bigskip
\begin{center}
{\Large {\bf{{{Phase spaces of twisted Lie-algebraically deformed
relativistic and nonrelativistic symmetries}}}}}
\end{center}
\bigskip
\bigskip
\bigskip
\begin{center}
{{\large ${\rm {Marcin\;Daszkiewicz}}$ }}
\end{center}
\bigskip
\begin{center}
\bigskip

{
${\rm{Institute\; of\; Theoretical\; Physics}}$}

{
${\rm{ University\; of\; Wroc{\l}aw\; pl.\; Maxa\; Borna\; 9,\;
50-206\; Wroc{\l}aw,\; Poland}}$}

{
${\rm{ e-mail:\; marcin@ift.uni.wroc.pl}}$}

\end{center}
\bigskip
\bigskip
\bigskip
\bigskip
\bigskip
\bigskip
\bigskip
\bigskip
\bigskip
\begin{abstract}
The twisted Lie-algebraically deformed relativistic and
nonrelativistic phase spaces are constructed with the use of
Heisenberg double procedure. The corresponding Heisenberg
uncertainty principles are discussed as well.
\end{abstract}
\bigskip
\bigskip
\bigskip
\bigskip
\eject

\section{{{Introduction}}}

Recently, there were found   formal arguments, based mainly on
Quantum Gravity \cite{grav1}, \cite{grav2} and String Theory models
\cite{string1}, \cite{string2}, indicating that space-time at
Planck-length should be noncommutative, i.e. it should have a
quantum nature. On the other side, the main reason for such
considerations follows from many phenomenological suggestions, which
state that relativistic space-time symmetries should be modified
(deformed) at Planck scale, while  the classical Poincare invariance
still remains valid at larger distances \cite{1a}-\cite{1d}.

It is well known that  a proper modification of the Poincare Hopf
algebra can be realized in the framework of Quantum Groups
\cite{twist}. Hence, in accordance with the Hopf-algebraic
classification  of all deformations of relativistic and
nonrelativistic symmetries (see \cite{zakrzewski}, \cite{kowclas}),
one can distinguish two
  kinds of quite interesting quantum spaces. First of them
corresponds to the well-known canonical type of noncommutativity

\begin{equation}
[\;{ x}_{\mu},{ x}_{\nu}\;] =
i\theta_{\mu\nu}\;,
\label{wielkaslawia}
\end{equation}
\\
with antisymmetric constant tensor $\theta^{\mu\nu}$. Its
relativistic and nonrelativistic Hopf-algebraic realizations have
been discovered  with the use of  twist procedure (see \cite{twist})
of classical Poincare \cite{3e}, \cite{3a} and Galilei
\cite{daszgali}, \cite{daszdual} Hopf structures respectively.

 The second class of mentioned deformations introduces
the Lie-algebraic type of space-time noncommutativity

\begin{equation}
[\;{ x}_{\mu},{ x}_{\nu}\;] = i\theta_{\mu\nu}^{\rho}x_{\rho}\;,
\label{wielkaslawia1}
\end{equation}
\\
with particularly chosen coefficients $\theta_{\mu\nu}^{\rho}$ being
constants.  It is represented by two Lie-algebraically deformed
Poincare Hopf algebras. First of them, so-called $\kappa$-Poincare
algebra \;${\mathcal U}_{\kappa}(\mathcal{P})$, has been proposed in
\cite{4a}, \cite{4b} as a result of contraction limit of q-deformed
anti-De-Sitter Hopf structure. It leads to the
 $\kappa$-Minkowski space-time \cite{kappammin}, \cite{kappammin1}
\begin{equation}
[\;{ x}_{0},{ x}_{i}\;] = \frac{i}{\kappa}{ x}_{i}\;\;\;,\;\;\;[\;{
x}_{i},{ x}_{j}\;] =0\;, \label{kappamin}
\end{equation}
with  mass-like deformation parameter $\kappa$. Besides, it also
gives a formal framework for such theoretical constructions as
Double Special Relativity (see e.g. \cite{dsr1}-\cite{dsr3}), which
postulates two observer-independent scales, of velocity, describing
the speed of light, and of mass, which can be identify with
$\kappa$-parameter - the fundamental Planck mass.

 The $\kappa$-deformed dual Poincare quantum group \;$\mathcal{P}_{\kappa}$  has been provided in
\cite{dualzakrzewski}, while the $\kappa$-deformed Galilei Hopf
algebra \;${\mathcal U}_{\kappa}(\mathcal{G})$ and the corresponding
dual quantum group \;$\mathcal{G}_{\kappa}$, have been discovered in
\cite{kappagali} and \cite{dualkappagali}  by nonrelativistic
contraction (see \cite{inonu}-\cite{azca2}) of their relativistic
counterparts.

The second type of deformation associated with  noncommutativity
(\ref{wielkaslawia1}) is generated (similar to the canonical
deformation (\ref{wielkaslawia})) by twist procedure \cite{twist}.
The corresponding Hopf algebras  have been proposed at relativistic
and nonrelativistic level in \cite{lie2} (see also \cite{lie1}) and
\cite{daszgali}, while their dual quantum groups - in \cite{lie2}
and \cite{daszdual} respectively.

The basic
 properties of the Lie-algebraically  twisted symmetries have been
investigated recently in a context of nonrelativistic particle
subjected to the external constant force  \cite{daszwal},
 and in the case of  harmonic oscillator  model \cite{oscy}. In particular,
there was demonstrated that such a kind of quantum space-time
produces additional acceleration, as well as  the velocity and
position-dependent forces, acting additionally
on a moving  particle.\\

In this article we introduce the relativistic and nonrelativistic
phase spaces corresponding to the twisted Lie-algebraically deformed
Hopf algebras \cite{lie1}, \cite{daszgali} and \cite{daszdual}. In
the case of relativistic symmetries, we use the so-called Heisenberg
double procedure \cite{twist}, \cite{pf}, which assumes  that the
momentum and position sectors of considered phase spaces can be
identified with the translation generators  of \;${\mathcal
U}_{\xi}(\mathcal{P})$ and \;$\mathcal{P}_{\xi}$ Hopf algebras
respectively; the cross-relations, i.e. the commutation relations
between momentum and position variables are given by dual parings of
corresponding generators. In the case of nonrelativistic symmetries
the corresponding phase spaces are obtained by proper
nonrelativistic contractions of their relativistic counterparts.

It should be noted that the deformed  phase space
\;$\chi_{\kappa}(\mathcal{P})$ has been already constructed in the
case of $\kappa$-Poincare algebra \cite{pf}, while its basic
physically implications have been investigated in the series of
papers \cite{pf}-\cite{camelia}. Particulary, there was studied its
role in a context of Quantum Gravity (see e.g. \cite{camelia}),
Doubly Special Relativity Theory \cite{jurek2}, \cite{blaut},
Statistical Physics\footnote{See e.g. deformed black body radiation
law \cite{alex}.} \cite{maggiore}-\cite{alex} and
Friedman-Robertson-Walker cosmological model  \cite{alex},
\cite{jurekevo}.

The main motivation for the present studies is twofold. First of
all, such a construction completed our knowledge about the whole
considered Hopf structure, i.e. about the Hopf algebra
\;$\mathcal{U}(\mathcal{A})$, its dual quantum group $\mathcal{A}$,
and the corresponding phase space $\chi(\mathcal{A})$. On the other
side, the recovered phase spaces give a background for the studies
on physical implications of the Lie-algebraically twisted Poincare
and Galilei Hopf algebras. Following the mentioned above
$\kappa$-Poincare program \cite{pf}-\cite{jurekevo} one can applied
the provided phase spaces to the (for example) Quantum Gravitational
or Cosmological
considerations respectively.\\

The paper is organized  as follows. In  second Section we recall
necessary  facts  concerning the twisted Lie-algebraically deformed
Poincare Hopf algebras and their dual quantum group \cite{lie1}.
Section three is devoted to the corresponding twisted phase spaces
provided  with the use of  Heisenberg double procedure. The proper
Heisenberg uncertainty principles are discussed in Section four. The
 nonrelativistic phase spaces (and the corresponding Heisenberg uncertainty
principles) are derived and discussed  in Section five.  The results
are summarized   in the last Section.

\section{Twisted Lie-algebraically deformed  Poincare Hopf algebra and its dual quantum group}

In this Section we recall the results of paper   \cite{lie2} (see
also \cite{lie1}) concerning the Lie-algebraically twisted  Poincare
Hopf algebra \;$\mathcal{U}_{\xi}(\mathcal{P})$ and its dual quantum
group $\mathcal{P}_\xi$, with mass-like deformation parameter $\xi$.
Both structures are described by the following Abelian r-matrix
\begin{equation}
{\rm r}_\xi= \frac{1}{2\xi}\zeta^\lambda\,P_{\lambda }\wedge
M_{\alpha \beta } \;\;;\;\;\;\alpha,\;\beta - {\rm
fixed}\;\;,\;\;\zeta^\lambda \;\; {\rm
denotes\;\;dimensionless\;\;fourvector}\;, \label{swiatowid}
\end{equation}
satisfying the classical Yang-Baxter equation \cite{twist}.\\
After twist procedure\footnote{Due to the formula (\ref{swiatowid})
the corresponding twist factor has the form $\mathcal{F}_{\xi}= \exp
\,\frac{i}{2\xi}( \zeta^\lambda\,P_{\lambda }\wedge M_{\alpha \beta
})$.} the algebraic sector of $~\mathcal{U}_{\xi}(\mathcal{P})$
algebra remains classical
\begin{eqnarray}
&&\left[ \,M_{\mu \nu },M_{\rho \sigma }\,\right] =i\left( \eta
_{\mu \sigma }\,M_{\nu \rho }-\eta _{\nu \sigma }\,M_{\mu \rho
}+\eta _{\nu \rho }M_{\mu
\sigma }-\eta _{\mu \rho }M_{\nu \sigma }\right) \;,  \label{swarzyca1} \\
&~~&  \cr &&\left[ \,M_{\mu \nu },P_{\rho }\,\right] =i\left( \eta
_{\nu \rho }\,P_{\mu }-\eta _{\mu \rho }\,P_{\nu }\right)
\;\;\;,\;\;\;\left[ \, P_{\mu },P_{\nu }\,\right] =0\;,
\label{ssswarzyca1}
\end{eqnarray}
while the coproduct becomes deformed
\begin{eqnarray}
 \Delta_\xi(P_\mu)&=&\Delta
_0(P_\mu)+(-i)^\gamma\sinh \left(
\frac{i^{\gamma}}{2\xi}\zeta^\lambda P_\lambda \right)\wedge
\left(\eta_{\alpha \mu}P_\beta -\eta_{\beta \mu}P_\alpha \right)\label{swiatowid100}\\
&+&(\cosh \left(\frac{i^{\gamma}}{2\xi}\zeta^\lambda  P_\lambda
\right)-1)\perp \left(\eta_{\alpha \alpha }\eta_{\alpha \mu}P_\alpha
+\eta_{\beta \beta }\eta_{\beta \mu}P_\beta \right)\;,
\notag  \\
&~~&  \cr \Delta_\xi(M_{\mu\nu})&=&\Delta_0(M_{\mu\nu})+M_{\alpha
\beta }\wedge \frac{1}{2\xi}\zeta^\lambda \left(\eta_{\mu \lambda
}P_\nu-\eta_{\nu \lambda }P_\mu\right)\nonumber\\
&+&i\left[M_{\mu\nu},M_{\alpha \beta }\right]\wedge
(-i)^{\gamma}\sinh\left(\frac{i^{\gamma}}{2\xi}\zeta^\lambda
P_\lambda \right) \nonumber \\
&+&\left[\left[%
M_{\mu\nu},M_{\alpha \beta }\right],M_{\alpha \beta
}\right]\perp(-1)^{1+\gamma}
(\cosh\left(\frac{i^{\gamma}}{2\xi}\zeta^\lambda  P_\lambda  \right)-1)  \label{swiatowid200} \\
&+&M_{\alpha \beta
}(-i)^\gamma\sinh\left(\frac{i^{\gamma}}{2\xi}\zeta^\lambda
P_\lambda \right)\perp
\frac{1}{2\xi}\zeta^\lambda \left(\psi_\lambda P_\alpha -\chi_\lambda P_\beta \right) \notag \\
&+&\frac{1}{2\xi}\zeta^\lambda \left(\psi_\lambda \eta_{\alpha
\alpha }P_\beta +\chi_\lambda \eta_{\beta \beta }P_\alpha
\right)\wedge M_{\alpha \beta
}(-1)^{1+\gamma}(\cosh\left(\frac{i^{\gamma}}{2\xi}\zeta^\lambda
P_\lambda \right)-1)\;,
  \notag
\end{eqnarray}
where $\Delta_0(a) = a\otimes 1 + 1\otimes a$, $a\wedge b=a\otimes
b-b\otimes a$, $a\perp b=a\otimes b+b\otimes a$, $\psi_\gamma
=\eta_{j \gamma }\eta_{l i}-\eta_{i \gamma }\eta_{lj},\; \chi_\gamma
=\eta_{j \gamma }\eta_{k i}-\eta_{i \gamma }\eta_{k j}$,
$\eta_{\mu\nu} = (-,+,+,+)$, and $\gamma =0$ when $M_{\alpha\beta}$
is a boost or $\gamma =1$ for a space
rotation.\\
 The dual quantum group $\mathcal{P}_\xi$
has been discovered  with  the use of FRT procedure \cite{frt}. In
terms of (dual) base $\{\,{\Lambda} _{\ \alpha }^{\beta },a^\mu\,\}$
it is given by the following algebraic sector
\begin{eqnarray}
&&\left[\,  {a}^{\mu }, {a}^{\nu }\,\right] =\frac{i}{\xi}\zeta
^{\nu } ( \delta _{\ \alpha }^{\mu } {a}_{\beta }-\delta _{\ \beta
}^{\mu } {a}_{\alpha }) + \frac{i}{\xi}\zeta^\mu( \delta _{\ \beta
}^{\nu } {a}_{\alpha }-\delta _{\ \alpha }^{\nu } {a}_{\beta
})\;\;\;,\;\;\; [ \, {\Lambda}
_{\,\nu }^{\mu }, {\Lambda} _{\,\tau }^{\rho }\,] =0\;,   \label{swarozyc1} \\
&&[ \, {a}^{\mu }, {\Lambda} _{\ \rho }^{\nu }\,]
=\frac{i}{\xi}\zeta ^{\lambda } {\Lambda} _{\ \lambda }^{\mu }( \eta
_{\beta \rho } {\Lambda} _{\ \alpha }^{\nu }-\eta _{\alpha \rho }
{\Lambda} _{\ \beta }^{\nu }) +\frac{i}{\xi}\zeta ^{\mu }( \delta
_{\ \beta }^{\nu } {\Lambda} _{\alpha \rho }-\delta _{\ \alpha
}^{\nu } {\Lambda} _{\beta \rho }) \;, \label{swarozyc2}
\end{eqnarray}
and the primitive coproducts
\begin{equation}
\Delta \,( {\Lambda} _{\ \nu }^{\mu })= {\Lambda} _{\ \rho }^{\mu
}\otimes  {\Lambda} _{\ \nu }^{\rho }\;\;\;,\;\;\; \Delta ( {a}^{\mu
})= {\Lambda} _{\ \nu }^{\mu }\otimes  {a}^{\nu }+ {a}^{\mu }\otimes
1\;.  \label{perun}
\end{equation}

\section{Relativistic phase spaces from Heisenberg double procedure}

The knowledge of Hopf algebra \;$\mathcal{U}_{\xi}(\mathcal{P})$ and
its dual quantum group \;$\mathcal{P}_\xi$ allows us to find the
corresponding $\xi$-deformed phase space
\;$\mathcal{\chi}_{\xi}(\mathcal{P})$. As it was mentioned in
Introduction, in accordance with Heisenberg double procedure
\cite{twist}, \cite{haidou}, the position sector of such a phase
space can be identified with translations ${a}^{\mu }$ (see
relations (\ref{swarozyc1})), while the momentum part -  with
generators $P_{\mu }$ (see formula (\ref{ssswarzyca1})). In order to
find the so-called cross-relations, i.e. the commutation relations
between positions and momenta one should use the  formula (see e.g.
\cite{haidou})
\begin{equation}
\left[\, \mathcal{Q} , \mathcal{R}\,\right] =
\mathcal{R}_{(1)}<\mathcal{Q}_{(1)},\mathcal{R}_{(2)}>\mathcal{Q}_{(2)}
- \mathcal{R}\mathcal{Q}\;, \label{perun1}
\end{equation}
where $<.,.>$ denotes  paring between generators $\mathcal{R}\in
\{\,M_{\mu \nu },P_{\rho }\,\}$ and $\mathcal{Q}\in \{\,{\Lambda}
_{\ \alpha }^{\beta },a^\mu\,\}$
\begin{eqnarray}
< \Lambda^{\mu}_{~\nu},1 > \,= \delta^{\mu}_{~\nu}\;\;\;,\;\;\; <
\Lambda^{\mu}_{~\nu},M^{\alpha\beta} > \,= i\left(\eta^{\alpha\mu}
\delta^{\beta}_{~\nu} - \eta^{\beta\mu} \delta^{\alpha}_{~\nu}
\right)\;\;\;,\;\;\; < a^{\mu},P_{\nu} > \,= i\delta^{\mu}_{~\nu}\;,
\label{rodnia}
\end{eqnarray}
and where we use Sweedler (shorthand) notation for coproduct $\Delta
(\mathcal{R}) = \sum \mathcal{R}_{(1)} \otimes \mathcal{R}_{(2)}$.

Let us start with "rotation-like"\footnote{Below, we consider three
kinds of twist factor (\ref{swiatowid}), providing different types
of space-time noncommutativity (see \cite{lie2},
\cite{lie1}).},\footnote{By "rotation-like" twist carrier we mean
the carrier containing space rotation generator $M_{kl}$.} twist
carrier $\{\,M_{kl}, P_\gamma\;;\; \gamma \ne\,k,\,l,\,0\,\}$
($\lambda = \gamma$, $\alpha = k$, $\beta = l\,$ in the formula
(\ref{swiatowid})). Then, in accordance with the above prescription,
we get the corresponding phase space (see (\ref{swiatowid100}),
(\ref{perun}) and (\ref{perun1}))\footnote{We put the nonzero
components of fourvector $\zeta$ equal one.}
\begin{eqnarray}
&{\bf i)}&\left[\, x_0, x_i\,\right] = \left[\, x_k, x_l\,\right]
=\left[\, p_\mu, p_\nu\,\right] =0\;\;\;;\;\;\;i=k,l,\gamma\;, \cr
&~~& \cr &&\left[\, x_k, x_\gamma\,\right] = \frac{i}{\xi}
x_l\;\;\;,\;\;\;\left[\, x_l, x_\gamma\,\right] = -\frac{i}{\xi} x_k
\;,\cr &~~&  \cr && \left[\, x_0, p_i\,\right] = \left[\, x_i,
p_0\,\right] = \left[\, x_k, p_\gamma\,\right] = \left[\, x_l,
p_\gamma\,\right] =0\;, \cr &~~& \cr && \left[\, x_0, p_0\,\right] =
-i\;\;\;,\;\;\;\left[\, x_\gamma, p_\gamma\,\right] =i
\;,\label{zerca}\\ &~~&  \cr &&\left[\, x_\gamma, p_k\,\right] =
\frac{i}{2\xi}p_l\;\;\;,\;\;\;\left[\, x_\gamma, p_l\,\right] =
-\frac{i}{2\xi}p_k\;,\cr&~~&  \cr && \left[\, x_l,
p_l\,\right]=i\cos\left(\frac{p_\gamma}{2\xi}\right)= \left[\, x_k,
p_k\,\right]\;,\cr &~~&  \cr && \left[\, x_k, p_l\,\right] =i \sin
\left(\frac{p_\gamma}{2\xi}\right)=-\left[\, x_l, p_k\,\right] \;.
\nonumber
\end{eqnarray}
In the case of carrier $\{\,M_{kl}, P_0\,\}$ we obtain
\begin{eqnarray}
&{\bf ii)}&\left[\, x_0, x_a\,\right] = \left[\, x_k, x_l\,\right]
=\left[\, p_\mu, p_\nu\,\right] =0\;\;\;;\;\;\;a \ne k,l,0\;, \cr
&~~& \cr &&\left[\, x_0, x_k\,\right] = \frac{i}{\xi}
x_l\;\;\;,\;\;\;\left[\, x_0, x_l\,\right] = -\frac{i}{\xi} x_k
\;\;\;,\;\;\;\left[\, x_k, x_a\,\right] = \left[\, x_l,
x_a\,\right]=0 \;,\cr &~~&  \cr && \left[\, x_0, p_a\,\right] =
\left[\, x_a, p_0\,\right] = \left[\, x_k, p_a\,\right] = \left[\,
x_l, p_a\,\right] = 0\;,\cr   &~~& \cr && \left[\, x_0, p_0\,\right]
= -i\;\;\;,\;\;\;\left[\, x_a, p_a\,\right] =i \;\;\;,\;\;\;\left[\,
x_a, p_k\,\right] =\left[\, x_a, p_l\,\right] = 0
\;,\nonumber\\
&~~& \cr &&\left[\, x_k, p_0\,\right] =0\;\;\;,\;\;\; \left[\, x_l,
p_0\,\right] =0\;,\label{zerca2}\\
&~~&  \cr &&\left[\, x_0, p_k\,\right] =
-\frac{i}{2\xi}p_l\;\;\;,\;\;\;\left[\, x_0, p_l\,\right] =
\frac{i}{2\xi}p_k\;,\cr&~~&  \cr && \left[\, x_l,
p_l\,\right]=i\cos\left(\frac{p_0}{2\xi}\right)= \left[\, x_k,
p_k\,\right] \;,\cr &~~&  \cr && \left[\, x_k, p_l\,\right] =i \sin
\left(\frac{p_0}{2\xi}\right)= -\left[\, x_l, p_k\,\right]\;,
\nonumber
\end{eqnarray}
while for "boost-like"\footnote{By "boost-like" twist carrier we
mean the carrier containing boost generator $M_{k0}$.} carrier
$\{\,M_{k0}, P_l\;;\; k\ne\,l\,\}$, we have
\begin{eqnarray}
&{\bf iii)}&\left[\, x_0, x_a\,\right] =\left[\, x_0, x_k\,\right] =
\left[\, p_\mu, p_\nu\,\right] =0\;\;\;;\;\;\;a \ne k,l,0\;, \cr
&~~&  \cr &&\left[\, x_k, x_a\,\right] = \left[\, x_l, x_a\,\right]
= 0 \;\;\;,\;\;\;\left[\, x_0, x_l\,\right] = \frac{i}{\xi}
x_k\;\;\;,\;\;\;\left[\, x_l, x_k\,\right] = -\frac{i}{\xi}
x_0\;,\cr &~~&  \cr && \left[\, x_l, p_k\,\right] =
\frac{i}{2\xi}p_0 \;\;\;,\;\;\; \left[\, x_0, p_0\,\right] =
-i\cosh\left(\frac{p_l}{2\xi}\right)\;,~~~~~~~~~~~\nonumber\\
&~~&  \cr && \left[\, x_a, p_a\,\right] = i \;\;\;,\;\;\; \left[\,
x_l, p_l\,\right] = i\;\;\;,\;\;\; \left[\, x_a, p_0\,\right] =
\left[\, x_k, p_l\,\right] =\left[\, x_0, p_l\,\right] =0\;,
\label{zerca3}\\ &~~& \cr &&\left[\, x_0, p_k\,\right] = i~{\rm
sinh}\left(\frac{p_l}{2\xi}\right) \;\;\;,\;\;\;\left[\, x_k,
p_k\,\right] =i\cosh\left(\frac{p_l}{2\xi}\right)\;, \nonumber\\
&~~& \cr &&\left[\, x_k, p_0\,\right] = -i~{\rm
sinh}\left(\frac{p_l}{2\xi}\right) \;\;\;,\;\;\;\left[\, x_l,
p_0\,\right] =\frac{i}{2\xi}p_k\;,\nonumber\\
&~~& \cr &&\left[\, x_k, p_a\,\right] =\left[\, x_l, p_a\,\right] =
\left[\, x_0, p_a\,\right] = \left[\, x_a, p_l\,\right] =\left[\,
x_a, p_k\,\right] = 0\;.\nonumber
\end{eqnarray}
The relation (\ref{zerca})-(\ref{zerca3}) describe three
relativistic phase spaces \;$\chi_{\xi}(\mathcal{P})$ associated
with the Lie-algebraically deformed Poincare Hopf algebra
\;$\mathcal{U}_{\xi}(\mathcal{P})$ and with  its (dual) quantum
group \;$\mathcal{P}_\xi$. Of course, for deformation parameter
$\xi$ running to infinity the above phase spaces become classical.
It should be also noted that for very particular choice of twist
factors (the choice of indices $\alpha$, $\beta$, $\gamma$) one can
recover the phase space proposed in \cite{phaseczer}.

\section{Heisenberg uncertainty principle}

Let us now turn to the Heisenberg uncertainty relations associated
with the above  phase spaces. 
If we
introduce the dispersion of observable ${\hat a}$ in a quantum
mechanical sense by (see e.g. \cite{camelia})\footnote{We put $\hbar
=1$.}
\begin{equation}
\Delta ({\hat a}) = \sqrt{<{\hat a}^2> - <{\hat
a}>^2}\;\;\;,\;\;\;\Delta ({\hat a})\Delta ({\hat b}) \geq
\frac{1}{2}|<{\hat c}>|\;, \label{slowianie}
\end{equation}
where ${\hat c} = [\, {\hat a},{\hat b} \,]$, than, we get the
following generalized (deformed) Heisenberg relations for all
considered above carriers: \\
\\
$\bullet$ for "rotation-like" carrier $\{\,M_{kl}, P_\gamma\;;\;
\gamma \ne\,k,\,l,\,0\,\}$
\begin{eqnarray}
&{\bf i)}& \Delta (x_k)\Delta (x_\gamma) \geq
\frac{|<x_l>|}{2\xi}\;\;\;,\;\;\;\Delta (x_l)\Delta (x_\gamma) \geq
\frac{|<x_k>|}{2\xi} \;,\cr &~~& \cr && \Delta (x_k)\Delta (p_k)
\geq \frac{|<\cos\left(\frac{p_\gamma}{2\xi}\right)>|}{2}
\;\;\;,\;\;\;\Delta (x_l)\Delta (p_l) \geq
\frac{|<\cos\left(\frac{p_\gamma}{2\xi}\right)>|}{2} \;,\cr &~~& \cr
&& \Delta (x_0)\Delta (p_0) \geq \frac{1}{2}\;\;\;,\;\;\;\Delta
(x_\gamma)\Delta (p_\gamma) \geq \frac{1}{2} \;,\label{slowianie1}\\
&~~&  \cr &&\Delta (x_\gamma)\Delta (p_k) \geq\frac{|<p_l>|}{4\xi}
\;\;\;,\;\;\;\Delta (x_\gamma)\Delta (p_l)
\geq\frac{|<p_k>|}{4\xi}\;, \cr&~~& \cr && \Delta (x_k)\Delta (p_l)
\geq \frac{|< \sin
\left(\frac{p_\gamma}{2\xi}\right)>|}{2}\;\;\;,\;\;\; \Delta
(x_l)\Delta (p_k) \geq \frac{|< \sin
\left(\frac{p_\gamma}{2\xi}\right)>|}{2} \;,\nonumber
\end{eqnarray}
$\bullet$ for twist carrier $\{\,M_{kl}, P_0\,\}$
\begin{eqnarray}
&{\bf ii)}& \Delta (x_k)\Delta (x_0) \geq
\frac{|<x_l>|}{2\xi}\;\;\;,\;\;\;\Delta (x_l)\Delta (x_0) \geq
\frac{|<x_k>|}{2\xi} \;,\cr &~~& \cr && \Delta (x_k)\Delta (p_k)
\geq \frac{|<\cos\left(\frac{p_0}{2\xi}\right)>|}{2}
\;\;\;,\;\;\;\Delta (x_l)\Delta (p_l) \geq
\frac{|<\cos\left(\frac{p_0}{2\xi}\right)>|}{2} \;,\cr &~~& \cr &&
\Delta (x_0)\Delta (p_0) \geq \frac{1}{2}\;\;\;,\;\;\;\Delta
(x_a)\Delta (p_a) \geq \frac{1}{2} \;,\label{slowianie2}\\
&~~&  \cr &&\Delta (x_0)\Delta (p_k) \geq\frac{|<p_l>|}{4\xi}
\;\;\;,\;\;\;\Delta (x_0)\Delta (p_l) \geq\frac{|<p_k>|}{4\xi}\;,
\cr&~~& \cr && \Delta (x_k)\Delta (p_l) \geq \frac{|< \sin
\left(\frac{p_0}{2\xi}\right)>|}{2} \;\;\;,\;\;\; \Delta (x_l)\Delta
(p_k) \geq \frac{|< \sin \left(\frac{p_0}{2\xi}\right)>|}{2}
\;,~~\nonumber
\end{eqnarray}
$\bullet$ and for "boost-like" carrier $\{\,M_{k0}, P_l\;;\;
k\ne\,l\,\}$
\begin{eqnarray}
&~{\bf iii)}& \Delta (x_k)\Delta (x_l) \geq
\frac{|<x_0>|}{2\xi}\;\;\;,\;\;\;\Delta (x_l)\Delta (x_0) \geq
\frac{|<x_k>|}{2\xi} \;,\cr &~~& \cr && \Delta (x_l)\Delta (p_k)
\geq \frac{|<p_0>|}{4\xi} \;\;\;,\;\;\;\Delta (x_0)\Delta (p_0) \geq
\frac{|<\cosh\left(\frac{p_l}{2\xi}\right)>|}{2} \;,\cr &~~& \cr &&
\Delta (x_l)\Delta (p_l) \geq \frac{1}{2}\;\;\;,\;\;\;\Delta
(x_a)\Delta (p_a) \geq \frac{1}{2}
\;,\label{slowianie3}\\
 &~~& \cr &&
\Delta (x_0)\Delta (p_k) \geq \frac{|<{\rm
sinh}\left(\frac{p_l}{2\xi}\right)>|}{2} \;\;\;,\;\;\;\Delta
(x_k)\Delta (p_k) \geq
\frac{|<\cosh\left(\frac{p_l}{2\xi}\right)>|}{2}
 \;,\nonumber\\
&~~& \cr && \Delta (x_k)\Delta (p_0) \geq \frac{|<{\rm
sinh}\left(\frac{p_l}{2\xi}\right)>|}{2} \;\;\;,\;\;\;\Delta
(x_l)\Delta (p_0) \geq \frac{|<p_k>|}{4\xi}
 \;,\nonumber
\end{eqnarray}
respectively.

 Obviously, for deformation parameter $\xi$
approaching  infinity the above relations become classical. It
should be also noted that for momentum variables $p_\gamma = p_0 =
p_k = 2\xi n\pi$ $(n=0,\pm 1, \pm 2, ...)$ all terms containing
"sinus/cosinus" and "$\sinh/\cosh$" functions disappear, i.e. the
deformation of Heisenberg uncertainty relations
(\ref{slowianie1})-(\ref{slowianie3}) becomes "minimal".

\section{Nonrelativistic phase spaces and Heisenberg uncertainty principle}

\subsection{Nonrelativistic phase spaces}

In this Section we  provide three nonrelativistic phase spaces (see
(\ref{coppy1400})-(\ref{coppy1600}))   with the use of contraction
procedures of their relativistic counterparts
(\ref{zerca})-(\ref{zerca3}). In a first step of our contraction
scheme, we introduce the following redefinition of the relativistic
phase space variables and the  deformation parameter $\xi$,
respectively
\begin{eqnarray}
&{\bf i)}& x_i = y_i\;\;\;,\;\;\;x_0 = ct\;\;\;,\;\;\;p_0 =
\frac{\pi_0}{c}\;\;\;,\;\;\;p_i = \pi_i  \;\;\;,\;\;\; \xi = \xi\;,
\label{niklot1}\\
&~~&  \cr &{\bf ii)}& x_i = y_i\;\;\;,\;\;\;x_0 = ct\;\;\;,\;\;\;p_0
=
\frac{\pi_0}{c}\;\;\;,\;\;\;p_i = \pi_i  \;\;\;,\;\;\; \xi = \frac{\hat{\xi}}{c}\;,\\
&~~&  \cr &{\bf iii)}& x_i = y_i\;\;\;,\;\;\;x_0 =
ct\;\;\;,\;\;\;p_0 = \frac{\pi_0}{c}\;\;\;,\;\;\;p_i = \pi_i
\;\;\;,\;\;\; \xi = c\bar{\xi}\;. \label{niklot3}
\end{eqnarray}
Next, in a second step,  we rewrite  the phase spaces
(\ref{zerca})-(\ref{zerca3}) in terms of $t$, $y_i$, $\pi_0$,
$\pi_i$ variables and deformation parameters $\xi$, $\hat{\xi}$,
$\bar{\xi}$, and we take the (nonrelativistic) limit $c\to \infty$.
In such a way we get the following Galilean phase spaces in the
first case (see "rotation-like" carrier)
\begin{eqnarray}
&{\bf i)}&\left[\, t, y_i\,\right] = \left[\, y_k, y_l\,\right]
=\left[\, \pi_\mu, \pi_\nu\,\right] =0\;\;\;;\;\;\;i=k,l,\gamma \;,
\cr &~~&  \cr &&\left[\, y_k, y_\gamma\,\right] = \frac{i}{\xi}
y_l\;\;\;,\;\;\;\left[\, y_l, y_\gamma\,\right] = -\frac{i}{\xi}
y_k\;,\cr &~~&  \cr && \left[\, t, \pi_i\,\right] = \left[\, y_i,
\pi_0\,\right] = \left[\, y_k, \pi_\gamma\,\right] = \left[\, y_l,
\pi_\gamma\,\right] =0\;,\cr
 &~~&
\cr && \left[\, t, \pi_0\,\right] = -i\;\;\;,\;\;\;\left[\,
y_\gamma, \pi_\gamma\,\right] =i \;,\label{coppy1400}\\ &~~&  \cr
&&\left[\, y_\gamma, \pi_k\,\right] =
\frac{i}{2\xi}\pi_l\;\;\;,\;\;\;\left[\, y_\gamma, \pi_l\,\right] =
-\frac{i}{2\xi}\pi_k\;,\cr&~~&  \cr && \left[\, y_l,
\pi_l\,\right]=i\cos\left(\frac{\pi_\gamma}{2\xi}\right)=\left[\,
y_k, \pi_k\,\right] \;,\cr&~~&  \cr && \left[\, y_k, \pi_l\,\right]
=-i \sin \left(\frac{\pi_\gamma}{2\xi}\right)=-\left[\, y_l,
\pi_k\,\right] \;, \nonumber
\end{eqnarray}
in the second case (corresponding to the twist carrier $\{\,M_{kl},
P_0\,\}$)
\begin{eqnarray}
~~~~~~~~~~~&{\bf ii)}&\left[\, t, y_a\,\right] = \left[\, y_k,
y_l\,\right] =\left[\, \pi_\mu, \pi_\nu\,\right] =0\;\;\;;\;\;\;a
\ne k,l,0\;, \cr &~~& \cr &&\left[\,t, y_k\,\right] = \frac{i}{{\hat
\xi}} y_l\;\;\;,\;\;\;\left[\,t, y_l\,\right] = -\frac{i}{{\hat
\xi}} y_k\;\;\;,\;\;\; \left[\, y_k, y_a\,\right] =\left[\, y_l,
y_a\,\right] =0 \;,\cr &~~&  \cr && \left[\, t, \pi_a\,\right] =
\left[\, y_a, \pi_0\,\right]= \left[\, y_k, \pi_a\,\right] =
\left[\, y_l, \pi_a\,\right]=0\;, \cr &~~& \cr && \left[\, t,
\pi_0\,\right] = -i\;\;\;,\;\;\;\left[\, y_a, \pi_a\,\right] =i
\;\;\;,\;\;\;\left[\,
y_a, \pi_k\,\right]=\left[\, y_a, \pi_l\,\right] = 0 \;,\nonumber\\
&~~&  \cr &&\left[\, y_k, \pi_0\,\right] =\left[\, y_l,
\pi_0\,\right]
=0\;,\label{coppy1500}\\
&~~&  \cr &&\left[\, t, \pi_k\,\right] = -\frac{i}{2{\hat
\xi}}\pi_l\;\;\;,\;\;\;\left[\, t, \pi_l\,\right] = \frac{i}{2{\hat
\xi}}\pi_k\;,\cr&~~&  \cr && \left[\, y_l, \pi_l\,\right]
=i\cos\left(\frac{\pi_0}{2{\hat \xi}}\right) =\left[\, y_k,
\pi_k\,\right]\;, \cr&~~&  \cr &&\left[\, y_k, \pi_l\,\right] =i
\sin \left(\frac{\pi_0}{2{\hat \xi}}\right) = -\left[\, y_l,
\pi_k\,\right] \;,\nonumber
\end{eqnarray}
and in the last case (for "boost-like" carrier $\{\,M_{k0}, P_l\;;\;
k\ne\,l\,\}$)
\begin{eqnarray}
&{\bf iii)}&\left[\, t, y_a\,\right] =\left[\, t, y_k\,\right] =
\left[\, \pi_\mu, \pi_\nu\,\right] =0\;\;\;;\;\;\;a \ne k,l,0\;, \cr
&~~&  \cr &&\left[\, y_k, y_a\,\right] = \left[\, y_l, y_a\,\right]
= 0 \;\;\;,\;\;\;\left[\, t, y_l\,\right] = 0\;\;\;,\;\;\;\left[\,
y_l, y_k\,\right] = -\frac{i}{\bar{\xi}} t\;,\cr &~~&  \cr &&
\left[\, y_l, \pi_k\,\right] = 0 \;\;\;,\;\;\; \left[\, t,
\pi_0\,\right] = -i\;\;\;,\;\;\; \left[\, t, \pi_k\,\right] =
0\;\;\;,\;\;\;\left[\, y_k, \pi_k\,\right] =i
\;,~~~~~~\label{coppy1600}\\
&~~&  \cr && \left[\, y_a, \pi_a\,\right] = i \;\;\;,\;\;\; \left[\,
y_l, \pi_l\,\right] = i\;\;\;,\;\;\; \left[\, y_k, \pi_l\,\right]
=\left[\, t, \pi_l\,\right] =0\;, \nonumber\\
&~~& \cr &&\left[\, y_k, \pi_a\,\right] =\left[\, y_l,
\pi_a\,\right] = \left[\, t, \pi_a\,\right] = \left[\, y_a,
\pi_l\,\right] =\left[\, y_a, \pi_k\,\right] = 0\;,\nonumber\\ &~~&
\cr && \left[\, y_a, \pi_0\,\right] = \left[\, y_k, \pi_0\,\right] =
\left[\, y_l, \pi_0\,\right] = 0\;,\nonumber
\end{eqnarray}
respectively.

 It should be noted that all above phase spaces can be get by direct
 application of Heisenberg double procedure as well. As it was
 mentioned in Introduction, the corresponding Hopf structures (i.e. the corresponding Galilei Hopf algebras
 \;$\mathcal{U}_{\cdot}(\mathcal{G})$ and their dual quantum groups
 \;$\mathcal{G}_{\cdot}$) have been obtained  in \cite{daszgali},
 \cite{daszdual} with the use of contractions of their relativistic counterparts
\;$\mathcal{U}_{\xi}(\mathcal{P})$ and \;$\mathcal{P}_{\xi}$. It
should be mentioned, however, that such a treatment is more
complicated technically than one used in presented Subsection.

\subsection{Heisenberg uncertainty principle}

Let us now consider the  Heisenberg uncertanity relations
corresponding to the nonrelativistic phase spaces
(\ref{coppy1400})-(\ref{coppy1600}). Using (\ref{slowianie}), one
can check that they take the form

\begin{eqnarray}
&{\bf i)}& \Delta (y_k)\Delta (y_\gamma) \geq
\frac{|<y_l>|}{2\xi}\;\;\;,\;\;\;\Delta (y_l)\Delta (y_\gamma) \geq
\frac{|<y_k>|}{2\xi} \;,\cr &~~& \cr && \Delta (y_k)\Delta (\pi_k)
\geq \frac{|<\cos\left(\frac{\pi_\gamma}{2\xi}\right)>|}{2}
\;\;\;,\;\;\;\Delta (y_l)\Delta (\pi_l) \geq
\frac{|<\cos\left(\frac{\pi_\gamma}{2\xi}\right)>|}{2} \;,\cr &~~&
\cr &&\Delta (t)\Delta (\pi_0) \geq \frac{1}{2}\;\;\;,\;\;\;\Delta
(y_\gamma)\Delta (\pi_\gamma) \geq \frac{1}{2} \;,\label{hai1400}\\
&~~~~~~&  \cr &&\Delta (y_\gamma)\Delta (\pi_k)
\geq\frac{|<\pi_l>|}{4\xi} \;\;\;,\;\;\;\Delta (y_\gamma)\Delta
(\pi_l) \geq\frac{|<\pi_k>|}{4\xi}\;, \cr&~~& \cr && \Delta
(y_k)\Delta (\pi_l) \geq \frac{|< \sin
\left(\frac{\pi_\gamma}{2\xi}\right)>|}{2} \;\;\;,\;\;\; \Delta
(y_l)\Delta (\pi_k) \geq \frac{|< \sin
\left(\frac{\pi_\gamma}{2\xi}\right)>|}{2} \;,\nonumber
\end{eqnarray}
for the first deformation
\begin{eqnarray}
&{\bf ii)}& \Delta (y_k)\Delta (t) \geq \frac{|<y_l>|}{2{\hat
\xi}}\;\;\;,\;\;\;\Delta (y_l)\Delta (t) \geq \frac{|<y_k>|}{2{\hat
\xi}} \;,\cr &~~~~~~& \cr && \Delta (y_k)\Delta (\pi_k) \geq
\frac{|<\cos\left(\frac{\pi_0}{2{\hat \xi}}\right)>|}{2}
\;\;\;,\;\;\;\Delta (y_l)\Delta (\pi_l) \geq
\frac{|<\cos\left(\frac{\pi_0}{2{\hat \xi}}\right)>|}{2} \;,\cr &~~&
\cr &&\Delta (t)\Delta (\pi_0) \geq \frac{1}{2}\;\;\;,\;\;\;\Delta
(y_a)\Delta (\pi_a) \geq \frac{1}{2} \;,\label{hai1500}\\
&~~&  \cr &&\Delta (t)\Delta (\pi_k) \geq\frac{|<\pi_l>|}{4{\hat
\xi}} \;\;\;,\;\;\;\Delta (t)\Delta (\pi_l)
\geq\frac{|<\pi_k>|}{4{\hat \xi}}\;, \cr&~~& \cr && \Delta
(y_k)\Delta (\pi_l) \geq \frac{|< \sin \left(\frac{\pi_0}{2{\hat
\xi}}\right)>|}{2} \;\;\;,\;\;\;\Delta (y_l)\Delta (\pi_k) \geq
\frac{|< \sin \left(\frac{\pi_0}{2{\hat \xi}}\right)>|}{2}
\;,\nonumber
\end{eqnarray}
in the second case, and
\begin{eqnarray}
&{\bf iii)}& \Delta (y_k)\Delta (y_l) \geq
\frac{|<t>|}{2\bar{\xi}}\;\;\;,\;\;\; \Delta (t)\Delta (\pi_0) \geq
\frac{1}{2} \;\;\;,\;\;\;\Delta (y_l)\Delta (\pi_l) \geq \frac{1}{2}
\;,\cr &~~& \cr && \Delta (y_a)\Delta (\pi_a) \geq \frac{1}{2}
\;\;\;,\;\;\;\Delta (y_k)\Delta (\pi_k) \geq
\frac{1}{2}\;,\label{hai1600}
\end{eqnarray}
for the last twist factor.

Of course, for  deformation parameters $\xi$, ${\hat \xi}$ and
${\bar \xi}$ approaching infinity the
 relations (\ref{coppy1400})-(\ref{coppy1600}) as well as (\ref{hai1400})-(\ref{hai1600})
become classical. Moreover, for momentum variables $\pi_\gamma =
\pi_0 = \pi_k = 2\alpha n\pi$ $(n=0,\pm 1, \pm 2, ...; \alpha = \xi,
\hat{\xi}, \bar{\xi})$ the oscillating and expanding terms in the
above relations disappear.

\section{{{Final Remarks}}}

In this article we construct six relativistic and nonrelativistic
phase spaces corresponding to the Lie-algebraically deformed
 Poincare and Galilei Hopf algebras respectively. The
considered phase spaces are provided with the use of Heisenberg
double procedure \cite{twist}.

It should be noted that presented results compleat our studies on
the  Lie-algebraically twisted groups at both levels - at the level
of relativistic and nonrelativistic symmetries as well. Moreover,
the provided phase spaces constitute the background for future
construction of basic dynamical models associated with  twisted
symmetries. As it was mentioned in Introduction, such investigations
have been already performed in the case of nonrelativistic particle
moving in a field of constant force \cite{daszwal}, and in the case
of harmonic oscillator model \cite{oscy}. However, used in
\cite{daszwal}, \cite{oscy} phase spaces have been taken "ad hoc",
i.e. without any formal, quantum group-like construction, such as,
for example, Heisenberg double procedure. The studies in this
direction are in progress.

\section*{Acknowledgments}
The author would like to thank J. Lukierski
for valuable discussions.\\
This paper has been financially supported by Polish Ministry of
Science and Higher Education grant NN202318534.

\end{document}